\documentclass[sigconf,natbib=true,anonymous=false]{acmart}
\usepackage{pifont}

\usepackage{multirow}
\usepackage{subfigure}
\usepackage{bm}
\usepackage{color}

\usepackage{enumitem}
\usepackage{graphicx}
\usepackage{mathtools}
\usepackage{colortbl}
\usepackage{balance}
\usepackage{graphics}
\usepackage{bbding}
\AtBeginDocument{%
  \providecommand\BibTeX{{%
    \normalfont B\kern-0.5em{\scshape i\kern-0.25em b}\kern-0.8em\TeX}}}

\copyrightyear{2023} 
\acmYear{2023} 
\setcopyright{acmlicensed}
\acmConference[CIKM '23]{Proceedings of the 32nd ACM International Conference on Information and Knowledge Management}{October 21--25, 2023}{Birmingham, United Kingdom} 
\acmBooktitle{Proceedings of the 32nd ACM International Conference on Information and Knowledge Management (CIKM '23), October 21--25, 2023, Birmingham, United Kingdom} 
\acmPrice{15.00} 
\acmDOI{10.1145/3583780.3614785} 
\acmISBN{979-8-4007-0124-5/23/10}

\begin{document}

\title{Attention Calibration for Transformer-based Sequential Recommendation}
\author{Peilin Zhou}
\authornote{Both authors contributed equally.}
\affiliation{%
  \institution{The Hong Kong University of Science and Technology (Guangzhou)}
  \country{}
  }
  \email{zhoupalin@gmail.com}

\author{Qichen Ye}
\authornotemark[1]
\affiliation{%
  \institution{Peking University}
  \country{}
  }
  \email{yeeeqichen@pku.edu.cn}

\author{Yueqi Xie}
\affiliation{%
  \institution{The Hong Kong University of Science and Technology}
  \country{}
  }
\email{yxieay@connect.ust.hk}

\author{Jingqi Gao}
\affiliation{%
  \institution{Upstage}
  \country{}
}
\email{mrgao.ary@gmail.com}

\author{Shoujin Wang}
\affiliation{%
  \institution{University of Technology Sydney}
  \country{}
  }
  \email{shoujin.wang@uts.edu.au}

\author{Jae Boum Kim}
\affiliation{%
  \institution{The Hong Kong University of Science and Technology}
  \country{}
  }
  \email{jbkim@cse.ust.hk}

\author{Chenyu You}
\affiliation{%
  \institution{Yale University}
  \country{}
  }
  \email{chenyu.you@yale.edu}

\author{Sunghun Kim}
\authornote{Corresponding author.}
\affiliation{%
  \institution{The Hong Kong University of Science and Technology (Guangzhou)}
  \country{}
  }
  \email{hunkim@cse.ust.hk}

\renewcommand{\shortauthors}{Peilin Zhou et al.}

\begin{abstract}
Transformer-based sequential recommendation (SR) has been booming in recent years, with the self-attention mechanism as its key component. Self-attention has been widely believed to be able to effectively select those informative and relevant items from a sequence of interacted items for next-item prediction via learning larger attention weights for these items. However, this may not always be true in reality. Our empirical analysis of some representative Transformer-based SR models reveals that it is not uncommon for large attention weights to be assigned to less relevant items, which can result in inaccurate recommendations. Through further in-depth analysis, we find two factors that may contribute to such inaccurate assignment of attention weights: \textit{sub-optimal position encoding} and \textit{noisy input}. To this end, in this paper, we aim to address this significant yet challenging gap in existing works. To be specific, we propose a simple yet effective framework called \textbf{A}ttention \textbf{C}alibration for \textbf{T}ransformer-based \textbf{S}equential \textbf{R}ecommendation (\textbf{AC-TSR}). In AC-TSR, a novel spatial calibrator and adversarial calibrator are designed respectively to directly calibrates those incorrectly assigned attention weights. The former is devised to explicitly capture the spatial relationships (i.e., order and distance) among items for more precise calculation of attention weights. The latter aims to redistribute the attention weights based on each item’s contribution to the next-item prediction. AC-TSR is readily adaptable and can be seamlessly integrated into various existing transformer-based SR models. Extensive experimental results on four benchmark real-world datasets demonstrate the superiority of our proposed AC-TSR via significant recommendation performance enhancements.
The source code is available at \url{https://github.com/AIM-SE/AC-TSR}.
\end{abstract}





\maketitle

\begin{figure}[t]
  \centering
  \includegraphics[width=70mm]{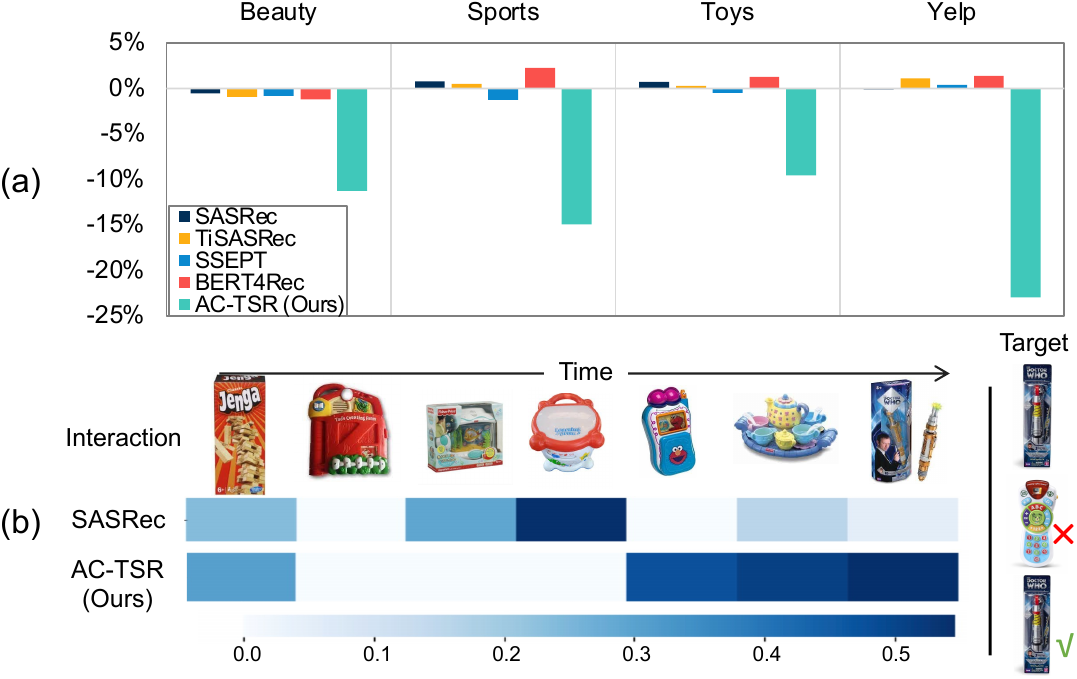}
  \caption{(a) Removing the highest attention weight from transformer-based SRS does not lead to a significant decrease in model performance and even improves performance in some cases; (b) Visualization of the attention weights from SASRec and our proposed AC-TSR.}
\label{fig:intro}
\vspace{-20pt}
\end{figure}

\section{Introduction}
\label{intro}
\par  
Sequential Recommender Systems (SRS) have been widely applied to various online services (\textit{e.g.}, e-commerce~\cite{hwangbo2018recommendation}, news websites~\cite{chu2019next}, social media~\cite{guy2010social}) to make recommendations on the next item which may interest a given user based on her/his historical interactions with items~\cite{wang2019sequential,lei2023practical}. In recent years, significant advancements have been made in the field of SRS by employing deep learning techniques~\cite{wang2020era,wang2021survey}. Various deep learning models, including recurrent neural networks (RNN)~\cite{hidasi2015session,song2021next}, convolutional neural networks (CNN)~\cite{tang2018personalized, yuan2019simple}, memory networks~\cite{chen2018sequential}, graph neural networks (GNN)~\cite{wu2019session, chang2021sequential}, and attention networks~\cite{wang2018attention} have been introduced to build a variety of SRS and have achieved great success. These SRS methods typically try to predict the next item of interest by well capturing the intricate sequential dependencies among items which have been sequentially interacted by users~\cite{wang2022exploiting,wang2022veracity}.     

In recent years, transformer-based SRS methods have gained significant attention. Benefiting from the particular self-attention mechanism, transformer-based methods are quite effective in capturing both short-term and long-term dependencies among items in a sequence. 
The pioneering work, SASRec~\cite{kang2018self}, introduced the self-attention mechanism to identify users' dynamic preferences based on their sequential interactions with items. BERT4Rec~\cite{sun2019bert4rec} extended SASRec to a bidirectional self-attentive architecture and incorporated a cloze task to capture contextual information from both the left and right sides of the target item.
Subsequent relevant studies have explored additional influential factors in the modeling process, such as time intervals~\cite{TiSASRec}, side information~\cite{Xie2022DIF}, and local constraints~\cite{he2021locker}, as well as techniques for reducing the model complexity~\cite{lightSANs, li2021lightweight, zhang2023efficiently}.
Self-attention mechanism, as the key component of transformer-based SRS methods, is thought to have the ability to accurately identify the influence of historical items on the next-item prediction by correctly assigning larger attention weights to those items which are more relevant to the target next item. However, our observations suggest that this may not always hold true in reality. In order to prove this statement, we conduct an empirical study that explicitly demonstrates our findings.

To be specific, we carefully carried out an "\textit{erasing}" experiment to examine the quality of the attention weights learned by the self-attention module in transformer-based SRS methods. In this experiment, we removed the highest attention weight learned by the self-attention module in various transformer-based SRS models and then examined the performance change caused by this removal. 
As shown in Fig.~\ref{fig:intro}(a), the removal operation only causes a marginal decrease in recommendation performance and even
improves performance in some cases, as indicated by the commonly used metric, Recall@20. This proved that the so-called most relevant and decisive item identified by the self-attention module in most transformer-based SRS actually has very little influence on the next-item prediction. This further indicates that self-attention mechanism may be deficient in identifying the decisive items within user behavior sequences. In order to further verify this statement, we visualized the attention weights learned by self-attention module in SASRec in Fig.~\ref{fig:intro}(b). Specifically, when given a sequence sample from the Amazon Toys dataset, which consists of a user's sequence of interacted items “\textit{Jenga, book, aquarium, drum, cell phone, tea set, laser screwdriver}” and the target next item “\textit{sonic screwdriver},” the self-attention mechanism of SASRec inaccurately assigns a larger attention weight to the item “\textit{drum},” which is totally irrelevant to the target next item. 

After careful and in-depth analysis, we found the aforementioned unreliable or inaccurate assignment of attention weights could be mainly attributed to the following two factors:  
(1) \textit{Sub-optimal position encoding.}
To capture the sequential dependencies over items, conventional transformer-based SRS methods directly incorporate item position embeddings into item embeddings and then use the integrated ones to compute the attention weights, suffering from the rank bottleneck~\cite{Xie2022DIF} and noisy correlations~\cite{lightSANs}. 
Recent studies \cite{Xie2022DIF,lightSANs} suggest decoupling the position encoding to calculate position correlations independently. 
However, such treatment is still sub-optimal since it fails to explicitly leverage low-level spatial information, including order and distance information, which have been found useful in enhancing the representation power of positional encoding in error-prone data~\cite{lee2022formnet}.
(2) \textit{Noisy input.}
Existing SRS works often assume a correlation between the target item and all historically interacted items.
However, in real-world scenarios, this assumption might not always hold since users may unconsciously interact with some items that deviate from their interests or preferences, resulting in noisy interaction data~\cite{sun2021does,wang2022trustworthy}.
Multiple factors, including users’ moods, social needs, personal conditions, etc., could lead to this noisy input phenomenon. For instance, users may randomly play popular videos or songs on a website that do not necessarily align with their preferences, or they might make purchases based on their mood or for their friends.
These sources of noise are challenging to identify as they often exhibit vague patterns, posing difficulties in accurately learning users' true preferences through the self-attention mechanism. 
Additionally, the self-attention mechanism is prone to overfitting on noisy input, further complicating the problem~\cite{zhou2022filter}.

To this end, in this paper, to address the aforementioned significant gaps in existing SRS studies, we propose a framework called \textbf{A}ttention \textbf{C}alibration for \textbf{T}ransformer-based \textbf{S}equential \textbf{R}ecommendation (\textbf{AC-TSR}). AC-TSR utilizes two well-designed calibrators, i.e., \textbf{Spatial Calibrator (SPC)} and \textbf{Adversarial Calibrator (ADC) }, to calibrate the unreliable attention weights illustrated above.
The SPC is designed to address the problem of sub-optimal position encoding by explicitly leveraging spatial relationships, such as the order and distance between items in a user sequence, to calculate attention weights that possess greater structural significance.
Specifically, SPC incorporates sequential relations directly into the attention matrix, eliminating the need for additional position embeddings such as absolute, relative, or decoupled ones.
The ADC tackles the issue of noisy input by redistributing attention weights based on the contribution of each historical item to the model's prediction. The term "adversarial" is employed because the reallocation process is performed in an adversarial manner.
The joint use of both calibrators can enhance the robustness of the original attention map against noisy input and enables more precise capture of user preferences.

Experiments on four benchmark datasets show that AC-TSR outperforms both non-transformer-based SRS methods and representative transformer-based SRS methods by calibrating the attention weights learned by the self-attention mechanism. 
Moreover, we demonstrate that both calibrators are plug-and-play and thus can be seamlessly incorporated into existing transformer-based SRS models to enhance their performance. 
Furthermore, we delve into each calibrator's influence and hyperparameters' effect in AC-TSR. A comprehensive analysis is carried out to shed light on why AC-TSR can achieve such performance gains.

In summary, we make the following contributions:
\begin{itemize}
    \item We propose the AC-TSR framework, which can effectively reduce the impact of sub-optimal position encoding and noisy input on the existing transformer-based SRS models with limited overhead.
    \item We develop two plug-and-play calibrators, namely spatial calibrator and adversarial calibrator, to rectify the unreliable attention. This ensures that the model focuses more on informative items when predicting the next item.
    \item We conduct comprehensive experiments on four benchmark datasets, demonstrating the superiority of AC-TSR over state-of-the-art SRS methods. 
\end{itemize}
\section{Related Work}
\subsection{Sequential Recommendation}
Sequential recommender systems (SRS) aim to predict the next item a user will likely interact with based on their past interactions. 
Early works, including pattern mining based~\cite{yap2012effective} and Markov chains~\cite{rendle2010factorizing, he2016fusing} based approaches focused on mining simple and low-order sequential dependencies, constrained by rigid assumptions and thus cannot deal with complex data with high order dependencies.
Later, with the advancements in neural networks, there has been a shift towards utilizing complex models such as Convolutional Neural Networks (CNNs)~\cite{tang2018personalized, yuan2019simple, yan2019cosrec}, Recurrent Neural Networks (RNNs)~\cite{hidasi2015session, ma2019hierarchical, peng2020ham, quadrana2017personalizing}, and Graph Neural Networks (GNNs)~\cite{chang2021sequential, wu2019session, guo2022evolutionary} to deal with the complex sequence patterns in the sequential recommendation task. 
For instance, Caser~\cite{tang2018personalized} leverages horizontal and vertical convolutional filters to discern sequential patterns hidden in the interaction data. 
GRU4Rec~\cite{hidasi2015session} utilizes a gated recurrent unit to analyze users' temporal behaviors. 
Furthermore, SR-GNN~\cite{wu2019session} transforms session sequences into graphs and applies graph neural networks to capture intricate item-item transitions.
 
More recently, transformer-based methods ~\cite{kang2018self, li2021lightweight, he2021locker, sun2019bert4rec, zhou2022equivariant, liu2023chatgpt,jiang2022adamct} have become the mainstream solutions due to their great potential in capturing user's sequential behavior through self-attention mechanism. 
SASRec~\cite{kang2018self} first adopts self-attention mechanism to capture users' sequential behaviors. 
And BERT4Rec~\cite{sun2019bert4rec} extends it to a bidirectional model with the help of Cloze task. 
In the follow-up studies, enhancements were made by incorporating time intervals~\cite{TiSASRec}, personalization~\cite{SSE-PT}, importance sampling~\cite{lian2020geography}, consistency~\cite{he2020consistency}, multiple interests~\cite{xie2023rethinking}, evolutionary preference~\cite{guo2022evolutionary}, continue learning~\cite{zhang2023survey, zhang2023continual} and decoupled positional encoding~\cite{lightSANs}. 
However, very few studies pay attention to the quality of the learned attention weights in transformer-based SRS.
In this study, we discover that in current transformer-based SRS models, the historical items with high attention weights do not consistently contribute to accurately predicting the target item.
Based on this observation, we propose to improve the quality of attention weights by injecting calibrators into transformer-based SRS models.
\vspace{-7.5pt}
\subsection{Debates on Attention Mechanism}
The debate surrounding the attention mechanism originates in the field of deep learning: for one thing, some researchers find that replacing high attention weights with lower ones does not affect the model's prediction performance~\cite{serrano2019attention,mohankumar2020towards,you2020contextualized,you2021mrd,you2022class}, possibly because the attention mechanism tends to assign higher weights to tokens that are not important, such as punctuation and stop words; for another, some studies in text classification~\cite{jain2019attention,chen2021adaptive} have revealed weak correlation between attention weights and gradient-based feature importance metrics. Furthermore, a recent machine translation work~\cite{lu2021attention} also observe that the attention mechanism fails to accurately identify the decisive inputs for each prediction, leading to incorrect or excessive translation in NMT. 
These debates on attention mechanisms inspire us to re-examine the learned attention weights in transformer-based SRS models.

In fact, there have been some SRS works that discuss the limitations of the attention mechanism:  Locker~\cite{he2021locker} contends that the self-attention modules struggle to capture the short-term user dynamics accurately. As a result the authors introduce an inductive local bias into the self-attention mechanism, improving the modeling of short-term user dynamics while maintaining long-term semantic information.
Rec-Denoiser~\cite{chen2022denoising} alleviates the influence of noise in the data by sparsifying the attention map, with the implicit assumption that not all attention weights carry important information in the self-attention layer. 
Different from these methods, our proposed AC-TSR effectively utilizes low-level spatial information through a spatial calibrator, and adopts an adversarial calibrator to adaptively identify the decisive items within the interaction sequence and automatically adjust the distribution of attention weights without relying on prior knowledge.
\begin{figure*}[t]
  \centering
  \includegraphics[width=165mm]{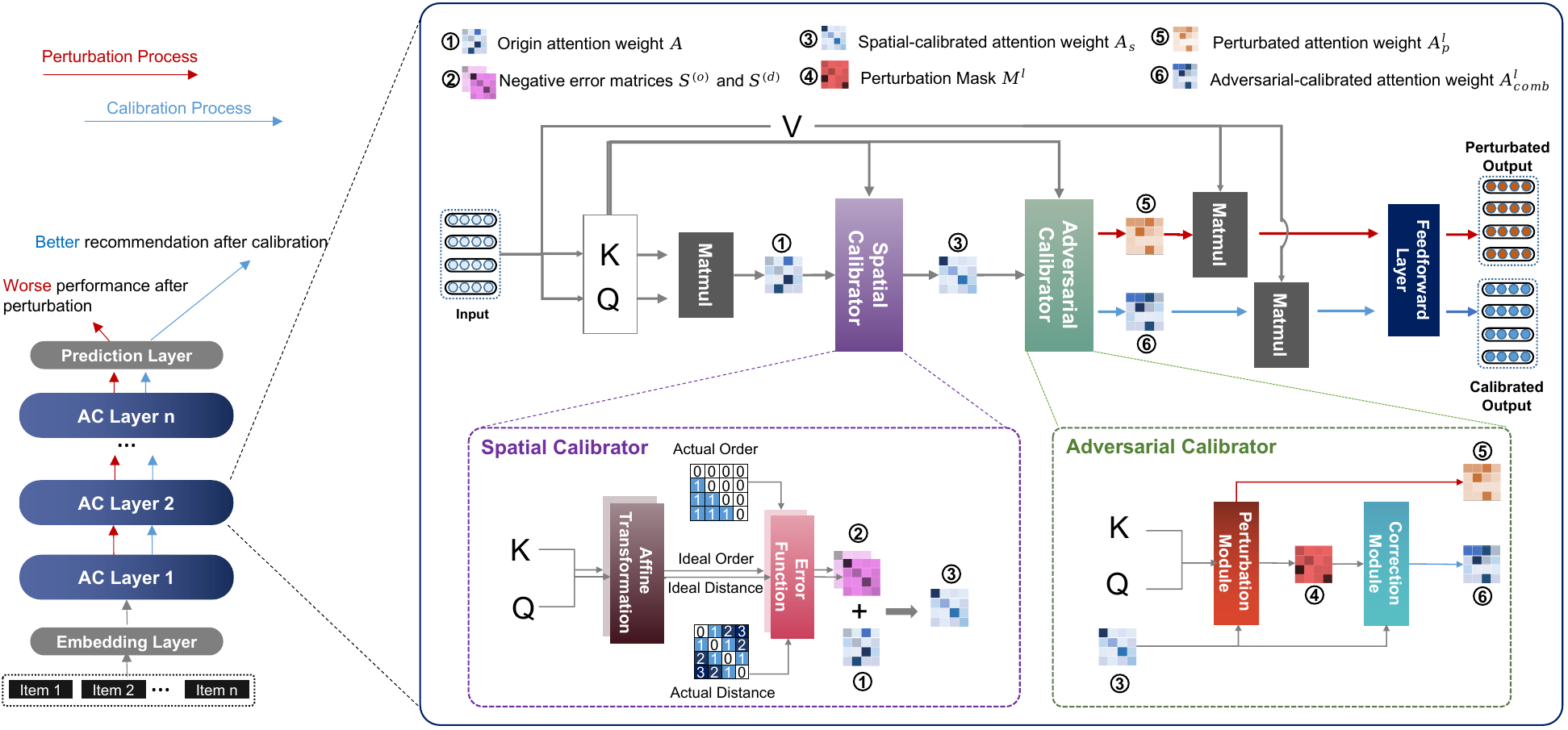}
  \caption{Overview of the proposed AC-TSR framework. The SASRec model functions as the backbone, where its self-attention layer is converted into an Attention Calibration (AC) layer for improved performance. Each AC layer contains a spatial calibrator (\textcolor[rgb]{0.41,0.29,0.54}{purple} dotted box) and an adversarial calibrator (\textcolor[rgb]{0.33,0.46,0.20}{green} dotted box). The spatial calibrator is responsible for incorporating spatial information such as order and distance into the attention weights. The adversarial calibrator aims to identify decisive items and adjust the distribution of attention weights.}
\label{fig:overview}
\end{figure*}

\section{Preliminary}
\subsection{Problem Setup}

Consider $\mathcal{U}$ as the set of users and $\mathcal{I}$ as the set of items.
For a user $u \in \mathcal{U}$, their past interactions can be represented as $\mathcal{S}^{u} = [v_1^u, v_2^u, \dots , v_{|\mathcal{S}^{u}|}^u]$. Here, $v^{u}_i \in \mathcal{I}$ represents the $i$-th interaction in the chronological sequence $\mathcal{S}^u$, with $|\mathcal{S}^{u}|$ indicating the sequence length. The set of users' actions is denoted as $\mathcal{S} = \{\mathcal{S}^1, \mathcal{S}^2, \dots, \mathcal{S}^{|\mathcal{U}|}\}$, where $|\mathcal{U}|$ is the total number of users.

The objective of sequential recommendation, given the historical interaction sequence $\mathcal{S}^u$ of user $u$, is to predict the next item $v_{\text{next}} \in \mathcal{I}$ that user $u$ will interact with at the $(|\mathcal{S}^{u}|+1)$-th time step, denoted as $p\left(v_{\text{next}} \mid \mathcal{S}^u\right)$.

\subsection{Transformer-based Recommenders}
Due to the remarkable capability in modeling sequential data, the transformer architecture has attracted a lot of attention and has been widely explored in the sequential recommendation~\cite{kang2018self,sun2019bert4rec,TiSASRec}.
Among these efforts, SASRec~\cite{kang2018self} is a particularly noteworthy piece of work, as it was the first transformer-based SR model and achieved competitive performance on many public datasets.
Therefore, in this session, we take SASRec as an example to offer a succinct overview of the transformer architecture.

\subsubsection{Embedding Layer}
\label{embedding}
The transformer-based recommenders maintain an item embedding table $\mathbf{T}\in \mathbb{R}^{|\mathcal{I}| \times d}$ to convert items from discrete ids to dense vectors, where $d$ represents the embedding size. 
First, a user interaction sequence $\mathcal{S}^u = [v_1^u, v_2^u, \dots , v_{|\mathcal{S}^{u}|}^u]$ is transformed into a fixed-length sequence $\hat{\mathcal{S}}^u = [v_1^u, v_2^u, \dots , v_{n}^u]$ by keeping most recent $n$ items or padding items, where $n$ is the maximum sequence length.
Then, the sequence embedding $\mathbf{E} \in \mathbb{R}^{n\times d}$ of $\hat{\mathcal{S}}^u$ is generated through the item embedding table $\mathbf{T}$. 
Finally, to consider the impact of different positions within the sequence, a learnable position embedding $\mathbf{P} \in \mathbb{R}^{n\times d}$ is added to $\mathbf{E}$ to obtain the final sequence embedding $\hat{\mathbf{E}}$, which serves as the input for the first transformer block.

\subsubsection{Transformer Block}
Transformer-based SR models typically employ multiple stacked transformer blocks to capture the hierarchical relationships among items within the input sequence.
Each transformer block comprises two main components: a self-attention layer and a point-wise feed-forward layer.

\noindent \textbf{Self-attention Layer:} The core of this layer is self-attention mechanism, which is designed to uncover the dependencies among items within a sequence~\cite{vaswani2017attention}. Specifically, the output item representation $\mathbf{H} \in \mathbb{R}^{n \times d}$ is calculated as follows:
\begin{equation}
\label{equation: self-attn}
    \mathbf{H} = \operatorname{Self-Attention}\left(\mathbf{Q}, \mathbf{K}, \mathbf{V}\right) = \operatorname{softmax}\left(\frac{\mathbf{Q}\mathbf{K}^T}{\sqrt{d}}\right)\mathbf{V},
\end{equation}
where $\mathbf{Q} = \hat{\mathbf{E}}\mathbf{W}_Q$, $\mathbf{K} = \hat{\mathbf{E}}\mathbf{W}_K$ and $\mathbf{V} = \hat{\mathbf{E}}\mathbf{W}_V$, $\left\{\mathbf{Q}, \mathbf{K}, \mathbf{V}\right\} \in \mathbb{R}^{n\times d}$ are the queries, keys and values respectively, and $\left\{\mathbf{W}_Q, \mathbf{W}_K, \mathbf{W}_V\right\} \in \mathbb{R}^{d \times d}$ are three learnable projection matrices, $\sqrt{d}$ is the scale factor. 
Thus, the attention weight $\mathbf{A} = \operatorname{softmax}\left(\frac{\mathbf{Q}\mathbf{K}^T}{\sqrt{d}}\right) $ establishes a whole-range connection between all items within sequence  $\hat{\mathcal{S}}^u$. 
Note that the Eq.\ref{equation: self-attn} can be further extended to the multi-head self-attention to obtain better expressiveness (\emph{i.e.} each attention head focuses on a different type of attention pattern~\cite{kovaleva2019revealing}).
Here we only show the single-head version for simplicity, and readers can check more details in the original paper~\cite{vaswani2017attention}. 

\noindent \textbf{Point-wise Feed-forward Layer:} This layer has the capability to learn complex feature representations, thereby enhancing the expressive power of the transformer architecture. 
After the calculation of self-attention, the output item representation $\mathbf{H}$ is fed to the point-wise feed-forward layer:
\begin{equation}
\label{equation: feed-forward}
    \mathbf{F}_i = \operatorname{FFN}\left(\mathbf{H}_i\right) = \operatorname{ReLU}\left(\mathbf{H}_i\mathbf{W}_1 + \mathbf{b}_1\right)\mathbf{W}_2 + \mathbf{b}_2,
\end{equation}
where $\mathbf{F}_i$ denotes the \emph{i-th} output embedding, $i \in [1, n]$; $\mathbf{W}_*$ and $\mathbf{b}_*$ are learnable weights and bias, respectively.
Note that in Eq.\ref{equation: feed-forward}, we leave out the residual connection, dropout, and layer normalization for simplicity.
In practice, these techniques can be adopted to enhance stability and speed up the training process. 

Transformer blocks are usually stacked to learn hierarchical item dependencies, and the output of $L$\textit{-th } block can be represented as:
\begin{equation}
    \mathbf{F}_i^L = \operatorname{FFN}\left(\mathbf{H}_i^L\right). 
\end{equation}

\subsubsection{Learning Objective}
In SASRec, the prediction of the next item $v_{next} \in \mathcal{I}$ is based on the last element of $\mathbf{F}^L$,  denoted as $\mathbf{F}^L_n$.
Specifically, the probability of interaction between user $u$ and each item is calculated through the inner product of $\mathbf{F}^L_n$ and the item embedding obtained from the item embedding table $\mathbf{T} \in \mathbb{R}^{|\mathcal{I}| \times d}$.
This process can be formulated as follows:
\begin{equation}
\label{equa:prediction}
    \hat{\mathbf{y}} = \operatorname{softmax}\left(\mathbf{T}{\mathbf{F}^L_n}^T\right),
\end{equation}
where $\hat{\mathbf{y}} \in \mathbb{R}^{|\mathcal{I}|}$ denotes the predicted probability. 
Afterwards, cross-entropy is chosen as the loss function to measure the discrepancy between the prediction $\hat{\mathbf{y}}$ and the ground truth $\mathbf{y}$:

\begin{equation}
\label{equa:loss}
    \mathcal{L} = -\sum^{|\mathcal{I}|}_{i=1}y_i\log\left(\hat{y}_i\right)
\end{equation}

\section{AC-TSR Framework}
\subsection{Overall Architecture}
The overall architecture of AC-TSR is depicted in Fig. \ref{fig:overview}, which introduces two new components, namely \textbf{Spatial Calibrator} (Sec.~\ref{sec: spatial-calibrator}) and \textbf{Adversarial Calibrator} (Sec.~\ref{sec: adversarial-calibrator}), into the self-attention layer of TSR. This modification transforms the original self-attention layer into an Attention Calibration (AC) layer. Inside each AC layer, the attention weights computed by the original attention mechanism are initially calibrated in sequence, starting with the Spatial Calibrator and followed by the Adversarial Calibrator.
Specifically, the Spatial Calibrator integrates spatial information, including order and distance, into the attention weights, while the Adversarial Calibrator aims to identify critical items and adjust the distribution of attention weights. Notably, the Adversarial Calibrator's output comprises both the perturbed attention weights and the calibrated attention weights for enhanced performance. These two weights are utilized to compute perturbed and calibrated representations, respectively. The perturbed representations are employed to disrupt the recommendation performance, whereas the calibrated representations are used to improve the recommendation performance. A more accurate attention distribution is ultimately obtained through the adversarial process of perturbation and calibration.
\subsection{Spatial Calibrator}
\label{sec: spatial-calibrator}
AC-TSR abandons traditional position encoding techniques, including absolute and relative embeddings, as they are hindered by the rank bottleneck~\cite{Xie2022DIF} and the noisy correlations~\cite{lightSANs}. 
Instead, it adopts a spatial calibrator (SPC) to empower the self-attention layer with the ability to recognize spatial relationships within the input sequence without the need for positional embeddings.
To this end, we first compute the \emph{order} and \emph{log-distance} between pairs of items with respect to the position in the input sequence and then directly use these low-level features to adjust the pre-softmax attention weights (\emph{i.e.,} the $\frac{\mathbf{Q}\mathbf{K}^T}{\sqrt{d}}$ in Eq.~\ref{equation: self-attn}). 
Specifically, the low-level features of actual \emph{order} $o_{ij}$ and actual \emph{log-distance} $d_{ij}$ between position $i$ and $j$ in the input sequence are defined as follows:
\begin{equation}
o_{ij} =\mathbb{I}(i<j)= \begin{cases}1, & i<j \\ 0, & otherwise\end{cases}
\end{equation}
\begin{equation}
    d_{ij} = \operatorname{ln}\left(1 + |i - j|\right).
\end{equation}
Then we use the query $\mathbf{q}^l_i$ and key $\mathbf{k}^l_j$ in each self-attention layer to predict the orders and distances the items should have if there exists a meaningful dependency between them:
\begin{equation}
\label{equa: spatial prediction 1}
    \hat{o}_{ij} = \operatorname{sigmoid}\left(\operatorname{affine}^{(o)}\left(\left[\mathbf{q}^l_i;\mathbf{k}^l_j\right]\right)\right),
\end{equation}
\begin{equation}
\label{equa: spatial prediction 2}
    \hat{d}_{ij} = \operatorname{affine}^{(d)}\left(\left[\mathbf{q}^l_i;\mathbf{k}^l_j\right]\right),
\end{equation}
where $\hat{o}_{ij}$ and $\hat{d}_{ij}$ denote predicted  order and distance, respectively. 
Finally, sigmoid cross-entropy and $L_2$ loss are adopted to calculate the discrepancies between the predictions and the ground truths, which are added into the origin attention weights:
\begin{equation}
\label{equa: spatial error 1}
    s_{ij}^{(o)} = o_{ij}\operatorname{ln}\left(\hat{o}_{ij}\right) + (1 - o_{ij})(1 - \operatorname{ln}\left(\hat{o}_{ij})\right),
\end{equation}
\begin{equation}
\label{equa: spatial error 2}
    s_{ij}^{(d)} = - \frac{\theta^2\left(d_{ij} - \hat{d}_{ij}\right)^2}{2},
\end{equation}
\begin{equation}
\label{equa: spatial attention}
    \mathbf{A}_{s} = \operatorname{softmax}\left(\frac{\mathbf{Q}\mathbf{K}^T}{\sqrt{d}} + \mathbf{s}^{(o)} + \mathbf{s}^{(d)}\right),
\end{equation}
where $\theta$ is a learnable scalar and $\mathbf{A}_{s}$ is the calibrated attention weights after spatial calibrator. 

The core idea of the spatial calibrator is to correct the attention weights by penalizing the attention edges that violate the order or distance constraints (\textit{i.e.} $o_{ij}$ or $d_{ij}$). 
Intuitively, if the self-attention mechanism can capture these low-level features, then we can assume that the spatial relationships are encoded in the query $\mathbf{Q}$ and key $\mathbf{K}$ since the attention weights are calculated based on them. 
In other words, the prediction errors calculated by Eq.~\ref{equa: spatial error 1} and Eq.~\ref{equa: spatial error 2} can reflect the potential weakness in the corresponding attention weights, so we calibrate it by adding a penalty to these positions.

\subsection{Adversarial Calibrator}
\label{sec: adversarial-calibrator}
The adversarial calibrator (ADC) aims to mitigate the noisy input issue mentioned in Sec.~\ref{intro} by making the self-attention mechanism more focused on the informative and decisive historical items.
To achieve this goal, we design a \textbf{Perturbation Module} and a \textbf{Correction Module} (\emph{cf.} Fig.~\ref{fig:overview}). 
Specifically, the perturbation module first automatically identifies the decisive historical items by adding limited perturbations to the original attention weights.
The correction module then calibrates the attention weights through highlighting the critical inputs (\emph{i.e.}, the perturbed positions) detected by the perturbation module. 
\subsubsection{Perturbation Module.}
The core idea of the perturbation module is to detect the decisive part in user sequence by perturbating original attention weights. 
For a transformer-based SR model, its performance is expected to be poor if the attention weights corresponding to the decisive parts are perturbed. 
Based on this, for the $l$-\textit{th} layer, a perturbation mask $\mathbf{M}^l$ is utilized to introduce uniform distribution $\mu$ to the spatial carlibrated attention weight $\mathbf{A}_{s}^l$, which simulates the process of perturbation: 

\begin{equation}
\label{equa: perturbed attention}
    \mathbf{A}_p^l = \mathbf{M}^l \odot \mathbf{A}_{s}^l + (1 - \mathbf{M}^l) \odot \mathbf{\mu },
\end{equation}
\begin{equation}
\label{equa: perturbation-mask}
    \mathbf{M}^l = \operatorname{sigmoid}\left(\frac{\mathbf{Q}^l\mathbf{W}_{Q_p}^l\left(\mathbf{K}^l\mathbf{W}_{K_p}^l\right)^T}{\sqrt{d}}\right),
\end{equation}  
where $\mathbf{A}_p^l$, $\mathbf{Q}^l$ and $\mathbf{K}^l$ are the perturbed attention weight, the query and the key in the $l$-th self-attention layer; $\odot$ denotes the element-wise multiplication; the  
$\mathbf{W}^l_{Q_p} \in \mathbb{R}^{d \times d}$ and $\mathbf{W}^l_{K_p} \in \mathbb{R}^{d \times d}$ are two learnable matrices.

\subsubsection{Correction Module.} As aforementioned, the perturbation module aims to deteriorate the model's performance by removing the decisive information through perturbation mask $\mathbf{M}^l$. 
In other words, $\mathbf{M}^l$ reveals the critical part in $\mathbf{A}_{s}^l$. 
To this end, we propose to adjust the spatial carlibrated attention weights $\mathbf{A}_{s}^l$ by highlighting the essential part:    
\begin{equation}
\label{equa: calibrated attention}
    \mathbf{A}_{c}^l = \mathbf{A}_{s}^l \odot e^{1 - \mathbf{M}^l}.
\end{equation}
In the above equation, we increase the attention weight in the spatial carlibrated attention $\mathbf{A}_{s}^l$ where the perturbation module module assigns large perturbation to (\emph{i.e.,} informative items that are critical for the model output). 
Afterwards, inspired by~\cite{Xie2022DIF}, we adopts a gating function to dynamically determine the amount of adversarial calibrated attention weight $\mathbf{A}_c^l$ to be fused into the spatial carlibrated attention weight $\mathbf{A}_{s}^l$, and the \textbf{combined} attention weight $\mathbf{A}_{comb}$ is calculated as follows:
\begin{equation}
    \mathbf{A}_{comb}^l = \mathbf{g} * \mathbf{A}_{s}^l + \left(1 - \mathbf{g}\right) * \mathbf{A}_c^l,
\end{equation}
\begin{equation}
    \mathbf{g} = \sigma\left(\mathbf{Q}^l\mathbf{W}_g^l + \mathbf{b}_g^l\right),
\end{equation}
where $\mathbf{W}_g^l \in \mathbb{R}^{d \times d}$, $\mathbf{b}_g^l \in \mathbb{R}^d$ are trainable parameters.

\subsection{Training Objective}
After obtaining the perturbed attention weight $\mathbf{A}_p$ and the combined attention weight $\mathbf{A}_{comb}$, we can further calculate the perturbed output embedding $\mathbf{\Tilde{F}}_p$ and calibrated output embedding $\mathbf{\Tilde{F}}_{c}$ by replacing the origin attention weight $\mathbf{A}$ in Eq.~\ref{equation: self-attn} with $\mathbf{A}_p$ and $\mathbf{A}_{comb}$, respectively. Then we can calculate the perturbed loss $\mathcal{L}_{P}$ and  calibrated loss $\mathcal{L}_{C}$ as follows:

\begin{equation}
\label{equa: perturbed output embedding}
    \mathcal{L}_{P} = -\sum^{|\mathcal{I}|}_{i=1}y_i\log\left(\hat{y}_i^{P}\right),
\end{equation}
\begin{equation}
\label{equa: calibrated output embedding}
    \mathcal{L}_{C} = -\sum^{|\mathcal{I}|}_{i=1}y_i\log\left(\hat{y}_i^{C}\right),
\end{equation}
where $\hat{y}_i^{P}$ and $\hat{y}_i^{C}$ are calculated by replacing $\mathbf{F}^L_n$ in Eq.~\ref{equa:prediction} with $\mathbf{\Tilde{F}}_{p, n}^L$ and $\mathbf{\Tilde{F}}_{c, n}^L$, respectively.

On the one hand, we want the model's performance based on the perturbed attention to be worse. On the other hand, we want to deteriorate the model's performance with as small perturbation as possible. 
Based on these two concerns, we define the final learning objective for the perturbation module as:
\begin{equation}
\label{equa:hyper-param}
    \mathcal{L}_{P_{final}}\left(\theta^{P}\right) = -\mathcal{L}_P\left(\theta\right) + \alpha \mathcal{L}_{norm} \left(\theta^{P}\right),
\end{equation}

\begin{equation}
    \mathcal{L}_{norm} \left(\theta^{P}\right) = \sum_{l=0}^L||1 - \mathbf{m}^l||_2,
\end{equation}
where $\theta^{P}$ denotes the parameters for perturbation module (\emph{i.e.,} $\left\{\mathbf{W}_{Q_p}, \mathbf{W}_{K_p}\right\}$ in all layers), and $\theta$ represents all other model parameters. $L$ is the number of transformer blocks and $\alpha$ is a hyper-parameter that balances $\mathcal{L}_P$ and $\mathcal{L}_{norm}$. 

Finally, we use the following loss to supervise our AC-TSR:
\begin{equation}
    \mathcal{L}_{final} = \mathcal{L}_{P_{final}} + \mathcal{L}_{C}.
\end{equation}
\subsection{Model Complexity}
\label{complexity}
The complexity of the AC-TSR model is derived from three components: the original transformer, the spatial calibrator, and the adversarial calibrator. The complexity of the original transformer remains consistent with that of backbone models such as SASRec and BERT4Rec. 
Inevitably, employing both calibrators simultaneously in the original transformer layer would increase the number of parameters and computational costs. 
To mitigate this issue, we propose several strategies for the training and inference stages to enhance computational efficiency:

\textbf{Training.} The two calibrators can be selectively applied to each transformer layer based on available computational resources, due to the effectiveness of using a single type of calibrator demonstrated by ablation study (Tab.~\ref{tab: ablation}).
For example, if temporal relationships are more significant in a dataset, one might prefer to only use the spatial calibrator. 
In extremely resource-constrained scenarios, one can opt to integrate a calibrator at just one layer rather than all layers. 
These strategies can effectively alleviate the computational burden introduced by the calibrators during the training stage, while keeping the time complexity in the same order as the original Transformer-based SR models.

\textbf{Inference.} We propose a lightweight version of AC-TSR, namely AC-TSR-lite, which excludes the two types of calibrators during the inference stage.
As a result, AC-TSR-lite maintains the same number of parameters and inference speed as the backbone TSR model. The performance of AC-TSR-lite will be elaborated in Sec~\ref{sec:overall_performance}.

\begin{table*}[t]
  \caption{Overall performance. The highest results are denoted in bold, while the runner-up results are underscored. 
  "*" indicates that the results are statistically significant with $p<0.01$ using a paired $t$-test.
}
  \label{tab:overall}
  \scalebox{0.70}{\setlength{\tabcolsep}{0.85mm}{
      \begin{tabular}{l|cccc|cccc|cccc|cccc}
        \toprule
        \multicolumn{1}{c|}{\multirow{3}{*}{SR Model}} & 
        \multicolumn{4}{c|}{\multirow{1}{*}{Beauty}} &
        \multicolumn{4}{c|}{\multirow{1}{*}{Sports}} &
        \multicolumn{4}{c|}{\multirow{1}{*}{Toys}} &
        \multicolumn{4}{c}{\multirow{1}{*}{Yelp}} 
          \\
          \cline{2-17}
          & 
        \multicolumn{2}{c}{\multirow{1}{*}{Recall}}&
        \multicolumn{2}{c|}{\multirow{1}{*}{NDCG}}
         &
        \multicolumn{2}{c}{\multirow{1}{*}{Recall}}&
        \multicolumn{2}{c|}{\multirow{1}{*}{NDCG}}
        &  
        \multicolumn{2}{c}{\multirow{1}{*}{Recall}}&
        \multicolumn{2}{c|}{\multirow{1}{*}{NDCG}} &
        \multicolumn{2}{c}{\multirow{1}{*}{Recall}}&
        \multicolumn{2}{c}{\multirow{1}{*}{NDCG}}
        \\
         &  
        @10 & @20 &
        @10 & @20 &
        @10 & @20 &
        @10 & @20 &
        @10 & @20 &
        @10 & @20 &
        @10 & @20 &
        @10 & @20 \\
        \midrule
PopRec & 0.0157 & 0.0242 & 0.0076 & 0.0097 & 0.0146 & 0.0244 & 0.0078 & 0.0103 & 0.0105 & 0.0172 & 0.0060 & 0.0077 & 0.0099 & 0.0161 & 0.0051 & 0.0067 \\
BPR & 0.0375 & 0.0590 & 0.0168 & 0.0222 & 0.0302 & 0.0480 & 0.0144 & 0.0188 & 0.0344 & 0.0560 & 0.0151 & 0.0205 & 0.0589 & 0.0830 & 0.0324 & 0.0384 \\
GRU4Rec & 0.0654 & 0.1002 & 0.0322 & 0.0410 & 0.0386 & 0.0609 & 0.0195 & 0.0251 & 0.0449 & 0.0708 & 0.0221 & 0.0287 & 0.0418 & 0.0679 & 0.0206 & 0.0271 \\
Caser & 0.0474 & 0.0731 & 0.0239 & 0.0304 & 0.0227 & 0.0364 & 0.0118 & 0.0153 & 0.0361 & 0.0566 & 0.0186 & 0.0238 & 0.0380 & 0.0608 & 0.0197 & 0.0255 \\
LightSANS & 0.0770 & 0.1177 & 0.0358 & 0.0461 & 0.0509 & 0.0781 & 0.0226 & 0.0294 & 0.0768 & 0.1116 & 0.0354 & 0.0442 & 0.0630 & 0.0904 & 0.0385 & 0.0453 \\
Locker & 0.0802 & 0.1197 & 0.0365 & 0.0464 & 0.0508 & 0.0753 & 0.0225 & 0.0286 & 0.0755 & 0.1094 & 0.0345 & 0.0430 & 0.0603 & 0.0869 & 0.0380 & 0.0446 \\ \hline
SASRec & 0.0779 & 0.1152 & 0.0353 & 0.0447 & 0.0504 & 0.0760 & 0.0224 & 0.0289 & 0.0776 & 0.1100 & 0.0352 & 0.0434 & 0.0618 & 0.0879 & 0.0387 & 0.0453 \\
\rowcolor{gray!10}\ w/ AC  & \underline{ 0.0817*} & \underline{ 0.1218*} & \textbf{0.0375*} & \textbf{0.0476*} & \underline{ 0.0532*} & \underline{ 0.0817*} & \underline{ 0.0235*} & \underline{ 0.0307*} & \underline{ 0.0825*} & \underline{ 0.1166*} & \underline{ 0.0371*} & \underline{ 0.0456*} & \textbf{0.0664*} & \textbf{0.0955*} & \textbf{0.0407*} & \textbf{0.0480*} \\
Improve. & 4.88\% & 5.73\% & 6.23\% & 6.49\% & 5.56\% & 7.50\% & 4.91\% & 6.23\% & 6.31\% & 6.00\% & 5.40\% & 5.07\% & 7.44\% & 8.65\% & 5.17\% & 5.96\% \\ \hline
BERT4Rec & 0.0557 & 0.0868 & 0.0279 & 0.0358 & 0.0313 & 0.0502 & 0.0155 & 0.0202 & 0.0489 & 0.0769 & 0.0253 & 0.0324 & 0.0467 & 0.0710 & 0.0264 & 0.0325 \\
\rowcolor{gray!10}\ w/ AC  & 0.0628* & 0.0929* & 0.0318* & 0.0394* & 0.0381* & 0.0607* & 0.0196* & 0.0253* & 0.0643* & 0.0924* & 0.0339* & 0.0410* & 0.0481* & 0.0769* & 0.0265* & 0.0337* \\
Improve. & 12.73\% & 7.03\% & 13.98\% & 10.06\% & 21.73\% & 20.92\% & 26.45\% & 25.25\% & 31.49\% & 20.16\% & 33.99\% & 26.54\% & 3.00\% & 8.31\% & 0.38\% & 3.69\% \\ \hline
SSE-PT & 0.0587 & 0.0936 & 0.0278 & 0.0366 & 0.0363 & 0.0580 & 0.0184 & 0.0239 & 0.0560 & 0.0837 & 0.0255 & 0.0325 & 0.0556 & 0.0779 & 0.0323 & 0.0379  \\
\rowcolor{gray!10}\ w/ AC  & 0.0629* & 0.1001* & 0.0293* & 0.0387* & 0.0379* & 0.0589* & 0.0191* & 0.0244* & 0.0614* & 0.0896* & 0.0282* & 0.0353* & 0.0565* & 0.0821* & 0.0330* & 0.0394* \\
Improve. & 7.16\% & 6.94\% & 5.40\% & 5.74\% & 4.41\% & 1.55\% & 3.80\% & 2.09\% & 9.64\% & 7.05\% & 10.59\% & 8.62\% & 1.62\% & 5.39\% & 2.17\% & 3.96\% \\ \hline
TiSASRec & 0.0794 & 0.1208 & 0.0356 & 0.0461 & 0.0523 & 0.0799 & 0.0230 & 0.0300 & 0.0819 & 0.1171 & 0.0367 & 0.0456 & 0.0618 & 0.0909 & 0.0387 & 0.0460 \\
\rowcolor{gray!10}\ w/ AC  & \textbf{0.0823*} & \textbf{0.1227*} & \underline{ 0.0373*} & \underline{ 0.0474*} & \textbf{0.0548*} & \textbf{0.0837*} & \textbf{0.0241*} & \textbf{0.0313*} & \textbf{0.0831*} & \textbf{0.1208*} & \textbf{0.0375*} & \textbf{0.0470*} & \underline{ 0.0654*} & \underline{ 0.0939*} & \underline{ 0.0401*} & \underline{ 0.0473*} \\
Improve. & 3.65\% & 1.57\% & 4.78\% & 2.82\% & 4.78\% & 4.76\% & 4.78\% & 4.33\% & 1.47\% & 3.16\% & 2.18\% & 3.07\% & 5.83\% & 3.30\% & 3.62\% & 2.83\% \\ 
      \bottomrule
    \end{tabular}
}
}
\end{table*}
\begin{table}[]
\caption{Model Complexity.}
\label{tab:model_complexity}
\resizebox{\columnwidth}{!}{%
\begin{tabular}{l|c|c|cccc}
\hline
\multirow{2}{*}{Model} & \multirow{2}{*}{\# Parameters} & \multirow{2}{*}{Inference speed} & \multicolumn{4}{c}{Recall@20} \\ \cline{4-7} 
 &  &  & Beauty & Sports & Toys & Yelp \\ \hline
SASRec & 0.87M & 2482.33/s & 0.1152 & 0.0760 & 0.1100 & 0.0879 \\
AC-SASRec & 0.90M & 917.54/s & 0.1218 & 0.0817 & 0.1166 & 0.0955 \\
AC-SASRec-lite & 0.87M & 2482.33/s & 0.1164 & 0.0768 & 0.1150 & 0.0913 \\ \hline
\end{tabular}%
}
\vspace{-10pt}
\end{table}
\section{Experiments}
In this study, we perform comprehensive experiments on four real-world datasets to address following research questions:
\begin{itemize}
    \item \textbf{RQ1:} Does the proposed AC-TSR exhibit competitive performance compared to current state-of-the-art transformer-based SR methods?
    \item \textbf{RQ2:} Can the Spatial Calibrator and Adversarial Calibrator be effectively integrated into representative transformer-based SR models and improve their performance?
    \item \textbf{RQ3:} What are the impacts of different components and hyper-parameters on AC-TSR's performance?
    \item \textbf{RQ4:} Why can AC-TSR achieve superior performance compared to other methods?
\end{itemize}

\subsection{Experimental Setup}
\label{sec:expe_setup}
\subsubsection{Datasets}
We carry out extensive experiments on a widely recognized dataset for business recommendations called \textbf{Yelp}~\footnote{https://www.yelp.com/dataset}, along with three categories of the Amazon Review dataset~\cite{DBLP:conf/sigir/McAuleyTSH15}: \textbf{Beauty}, \textbf{Sports}, and \textbf{Toys}.~\footnote{http://jmcauley.ucsd.edu/data/amazon/} We consider every interaction in these datasets as implicit feedback. For the Yelp dataset, we follow the preprocessing step described in~\cite{zhou2020s3}, keeping only the transactions after January 1st, 2019. Consistent with earlier research~\cite{kang2018self,zhou2020s3,yuan2021icai}, we exclude any items and users appearing less than five times in these datasets. The processed data's details across all four datasets can be found in \cite{Xie2022DIF}.

\subsubsection{Evaluation Metrics.}
For model evaluation, we employ the \textit{leave-one-out} strategy: we arrange each user-item interaction sequence chronologically, utilizing the latest interaction item as the test data, the second latest item as the validation data, and the remaining data as the training data. Consistent with existing sequential recommendation research~\cite{kang2018self,sun2019bert4rec}, we utilize two classical evaluation metrics, namely Recall@$K$ and NDCG@$K$, where NDCG represents Normalized Discounted Cumulative Gain, with $K$ taking values of 10 and 20.
Following the recommendation put forth by~\cite{krichene2020sampled, dallmann2021case}, we assess model performance using a full ranking approach, where the rankings are calculated across the entire set of items rather than just a subset.
\subsubsection{Baseline Methods}
\label{baselines}
We compare our proposed AC-TSR with 10 different baselines, including 2 general methods: \textbf{PopRec} and \textbf{BPR}~\cite{bpr}, 2 basic SR methods: \textbf{GRU4Rec}~\cite{hidasi2015session} and \textbf{Caser}~\cite{tang2018personalized}, and 6 representative transformer-based approaches: \textbf{LightSANS}~\cite{lightSANs},  \textbf{Locker}~\cite{he2021locker}, \textbf{BERT4Rec}~\cite{sun2019bert4rec}, \textbf{SASRec}~\cite{kang2018self}, \textbf{SSE-PT}~\cite{SSE-PT} and \textbf{TiSASRec}~\cite{TiSASRec}.
We do not select Rec-Denoiser~\cite{chen2022denoising} as a baseline because the authors have not released their codes and our own implementation is unsuccessful due to an out-of-memory error when computing the Jacobian matrix.
In addition, it is worth mentioning that there have been numerous transformer-based SR models introduced in recent years. However, we have not considered all of them as baselines due to the fact that some of these methods are not suitable for direct comparison due to their reliance on different training paradigms.

\subsubsection{Implementation Details}
\label{sec: implementation details}
For fair comparisions, we implement both the baseline methods and our proposed AC-TSR using the widely-used recommendation framework RecBole~\cite{zhao2021recbole} and evaluate them under identical settings. All baselines and AC-TSR are trained using the Adam optimizer for 200 epochs, employing a batch size of 256 and a learning rate of 1e-4.
The max sequence length of Sports, Toys, Beauty, and Yelp is set to 50. 
For LOCKER, we use a CNN as the local encoder (same as \emph{LOCKER+Conv} in their official implementation\footnote{https://github.com/AaronHeee/LOCKER}). For our proposed AC-TSR and other transformer-based baselines (\emph{e.g.}, SASRec), we perform grid search on other hyper-parameters to find the best combination. The searching space is: number of self-attention layers $\in \left\{2, 3\right\}$, number of self-attention heads $\in\left\{2, 4\right\}$, hidden size $\in \left\{64, 128\right\}$ and inner size $\in \left\{64, 128\right\}$. Both baselines and our method are carefully tuned on the used datasets for best performance. 

\subsection{Overall Performance (RQ1\&2)}
\label{sec:overall_performance}
The results of various methods across all datasets are summarized in Tab.~\ref{tab:overall}. Our proposed AC-TSR method exhibits superior performance on all datasets. Other key observations are as follows: 

\textbf{Most transformer-based methods, such as SASRec and TiSASRec, consistently outperform non-transformer-based methods like Caser and GRU4Rec by a large
margin. } This observation highlights the superiority of transformer-based methods in capturing long-range item dependencies in sequential data. 
Moreover, we observe that BERT4Rec falls to beat SASRec in the full-ranking evaluation setting, a fact also mentioned in prior studies~\cite{li2021lightweight, Xie2022DIF}.
One potential reason is that the original BERT4Rec paper employs a popularity-based sampling strategy for model evaluation. This strategy benefits BERT4Rec since its bi-directional encoder, combined with the Cloze task, allows for improved learning of representations for popular items.

\textbf{Auxiliary information can enhance the performance of transformer-based models.} 
By comparing the performance of SASRec, TiSASRec, and SSE-PT, we observe that transformer-based models can benefit from the integration of auxiliary information such as time intervals and user personality. This is due to auxiliary information providing essential cues for a more profound comprehension of users' dynamic behaviors. Capturing these behaviors accurately would be otherwise difficult, relying solely on the standard self-attention mechanism.

\textbf{Spatial Calibrator and Adversarial Calibrator can be seamlessly incorporated into existing transformer-based SR models (TSR) and boost their performance.} In our experiments, we choose SASRec, BERT4Rec, TiSASRec, and SSE-PT as backbones, which represent the unidirectional model, bidirectional model, and auxiliary information enhanced model (\emph{i.e.,} time interval and user embedding), respectively. 
And we adapt them into our proposed AC-TSR framework by dropping their position encoding module and replacing their transformer layer with our proposed Attention Calibration layer.
As shown in Tab.~\ref{tab:overall}, the TSR models integrated into the AC-TSR framework (highlighted in gray) exhibit significant performance improvements compared to the original TSR model. Specifically, AC-BERT4Rec achieves an average relative improvement of \textbf{17.24\%} and \textbf{18.7\%} in terms of Recall@10 and NDCG@10, respectively, across the four datasets. Similarly, AC-SASRec demonstrates average relative improvements of \textbf{6.05\%} and \textbf{5.43\%} for Recall@10 and NDCG@10, respectively. 
Comparable results are also observed in AC-TiSASRec and AC-SSE-PT.
These results demonstrate the effectiveness of our proposed method and its potential as a plug-in module for state-of-the-art transformer-based recommendation models.

In Sec.~\ref{complexity}, we discussed the impact of incorporating two calibrators into transformer layers, which inevitably increases computational cost and reduces inference speed. This presents a challenge for implementing AC-TSR in industrial scenarios.
To mitigate this issue, we propose a lightweight version of AC-TSR, called AC-TSR-lite. The key distinction between the lite version and the original AC-TSR lies in their usage of calibrators. In AC-TSR-lite, calibrators are employed only during the model training phase and removed during the inference phase, thereby maintaining the same structure as TSR.
To illustrate this, we compare the performance of SASRec, AC-SASRec, and AC-SASRec-lite on four datasets in Tab.\ref{tab:model_complexity}. We observe that in most cases, although AC-SASRec-lite exhibits lower Recall@20 compared to AC-SASRec, it still consistently outperforms SASRec and retains the same inference speed as SASRec. This provides evidence supporting the feasibility and flexibility of deploying AC-TSR in an industrial setting.

\begin{table}[t]
\centering
\caption{Ablation study of AC-TSR on Beauty dataset.}
\resizebox{65mm}{!}{
\begin{tabular}{l|c c|c|cccc}
\toprule
\multirow{2}*{\textbf{Settings}} & \multicolumn{2}{c|}{\textbf{Spatial}}& \multirow{2}*{\textbf{Adv.}} & \multicolumn{2}{c}{\textbf{Recall}} & \multicolumn{2}{c}{\textbf{NDCG}}\\
~ & \ \ order & distance & & @10 & @20 & @10 & @20 \\ 
\midrule
\quad (A) & {\color{green}\CheckmarkBold} & {\color{green}\CheckmarkBold} & {\color{green}\CheckmarkBold} & 0.0817 & 0.1218 & 0.0375 & 0.0476\\

\quad (B) & {\color{green}\CheckmarkBold} & {\color{red}\XSolidBrush} & {\color{green}\CheckmarkBold} & 0.0791 & 0.1210 & 0.0364 & 0.0470\\

\quad (C) & {\color{red}\XSolidBrush} & {\color{green}\CheckmarkBold} & {\color{green}\CheckmarkBold}
& 0.0792 & 0.1201 & 0.0372 & 0.0475\\

\quad (D) &  {\color{red}\XSolidBrush} &  {\color{red}\XSolidBrush} & {\color{green}\CheckmarkBold} & 0.0800 &0.1202 & 0.0367& 0.0469\\

\quad (E) & {\color{green}\CheckmarkBold} & {\color{green}\CheckmarkBold} &  {\color{red}\XSolidBrush} &0.0802 & 0.1197& 0.0365& 0.0464\\

\quad (F) & {\color{green}\CheckmarkBold} & {\color{red}\XSolidBrush} &  {\color{red}\XSolidBrush} & 0.0776 & 0.1168 & 0.0353 & 0.0452 \\

\quad (G) & {\color{red}\XSolidBrush} & {\color{green}\CheckmarkBold} &  {\color{red}\XSolidBrush} &0.0806 & 0.1188 & 0.0367 & 0.0463\\

\quad (H) & {\color{red}\XSolidBrush} & {\color{red}\XSolidBrush} &  {\color{red}\XSolidBrush} &0.0779  & 0.1152 & 0.0353 & 0.0447\\
\bottomrule
\end{tabular}}
\label{tab: ablation}
\vspace{-1.0em}
\end{table}

\begin{table}[t]
\centering
\caption{Impact of different positional encoding strategies. The SASRec is chosen as the backbone.}
\label{tab:position_ablation}
\resizebox{70mm}{!}{%
\begin{tabular}{l|cc|cc}
\hline
\multirow{2}{*}{\textbf{Position Encoding Strategy}} & \multicolumn{2}{c|}{\textbf{Sports}} & \multicolumn{2}{c}{\textbf{Toys}} \\
 & Recall@20 & NDCG@20 & Recall@20 & NDCG@20 \\ \hline
Remove Position & 0.0775 & 0.0294 & 0.1170 & 0.0456 \\
Absolute Position & 0.0760 & 0.0289 & 0.1100 & 0.0434 \\
Relative Position & 0.0753 & 0.0285 & 0.1172 & 0.0461 \\
Decoupled Position & 0.0769 & 0.0295 & 0.1153 & 0.0449 \\
Spatial Calibrator (Ours) & \textbf{0.0785} & \textbf{0.0298} & \textbf{0.1193} & \textbf{0.0462} \\ \hline
\end{tabular}%
}
\end{table}
\begin{figure}[t]
  \centering
  \includegraphics[width=70mm]{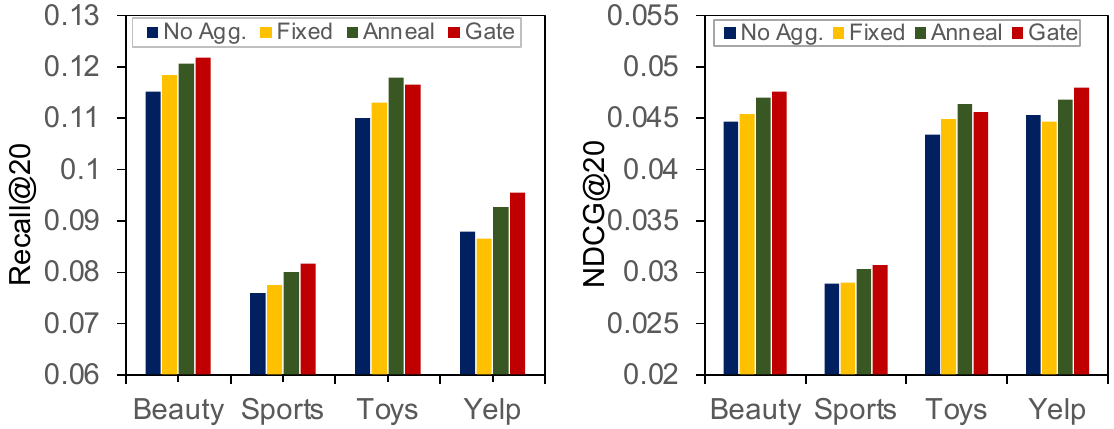}
  \caption{Impact of different aggregation strategies in Correction Module.}
\label{fig:aggregation}
\end{figure}
\begin{figure}[t]
  \centering
  \includegraphics[width=70mm]{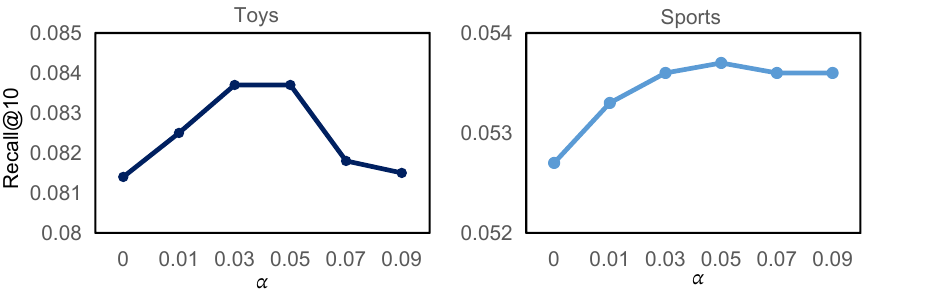}
  \caption{Effect of balance parameter $\alpha$.}
\label{fig:hyper-param}
\vspace{-1.5em}
\end{figure}

\subsection{Ablation and Hyper-parameter Studies (RQ3)}
\label{sec:in_depth_analysis}
\subsubsection{Contribution of Different Components.}
As shown in Table~\ref{tab: ablation}, we investigate 8 settings from a combination of (1) Whether to include order information in the Spatial Calibrator, (2) Whether to include distance information in the Spatial Calibrator, and (3) Whether to incorporate the Adversarial Calibrator. We can draw the following conclusions: First, by comparing the results of \emph{setting (F)} and \emph{setting (G)} with \emph{setting (H)}, which represents the initial SASRec, we can see that both the inclusion of order information and distance information in the Spatial Calibrator improve the model's performance. Combining order information and distance information together leads to even better performance compared to using them separately, thus validating the rationale behind our proposed Spatial Calibrator. Second, models equipped with our proposed Adversarial Calibrator consistently outperform those without it, while keeping other factors constant. For instance, we can compare the performance between \emph{setting (A)} and \emph{setting (E)}, or between \emph{setting (B)} and \emph{setting (F)}. This observation confirms that our Adversarial Calibrator effectively adjusts the attention weights in transformer-based SR models, resulting in improved performance.
\subsubsection{Comparison Between Spatial Calibrator and Different Positional Encoding Strategies.}
We test 4 representative position embedding approaches and our proposed spatial calibrator using SASRec as the backbone.
Due to limited space, we only present the comparison results on the Sports and Toys datasets in Tab.~\ref{tab:position_ablation}. 
To our surprise, we find that removing the position embedding from SASRec does not compromise the recommendation performance, but rather outperforms some methods using position embedding. 
For instance, on the Sports dataset, removing the position embedding achieves higher Recall@20 than the three methods using position embedding.
On the Toys dataset, both relative position embedding and removing positions achieved competitive performance. 
 Nonetheless, our proposed spatial calibrator still outperforms all compared position encoding strategies.
 We attribute this to the low-level features, including order and distance, being more effectively utilized by the self-attention mechanism, especially in noisy input scenarios.
\subsubsection{Comparison Between Gating Mechanism and Other Aggregation Strategies.}
We also investigate the impact of various attention aggregation strategies used in the correction module. In particular, we compare the gating mechanism with two commonly employed aggregation strategies: summation~\cite{Xie2022DIF} and annealing learning~\cite{lu2021attention}. The comparison results are presented in Fig.~\ref{fig:aggregation}. We can observe that the gating mechanism consistently outperforms the other two strategies in most cases, which can be attributed to its ability to dynamically control the proportions of original and calibrated attention weights. As a result, we ultimately chose the gating mechanism as the attention aggregation strategy in the correction module.

\subsubsection{Hyper-parameter Studies}
In this experiment, we aim to study the impact of hyper-parameter $\alpha$ in Eq.~\ref{equa:hyper-param}. Fig.\ref{fig:hyper-param} presents the Recall@10 and NDCG@10 scores of AC-TSR with different $\alpha$ on Toys and Sports datasets. From the figures, we can observe that the best performance is achieved when the balance parameter $\alpha$ is set to 0.03 or 0.05. Moreover, we observe that the impact of the balance parameter $\alpha$ varies across different datasets. Specifically, the performance fluctuates within a narrower range on the Sports dataset compared to the Toys dataset.
\begin{figure}[t]
  \centering
  \includegraphics[width=70mm]{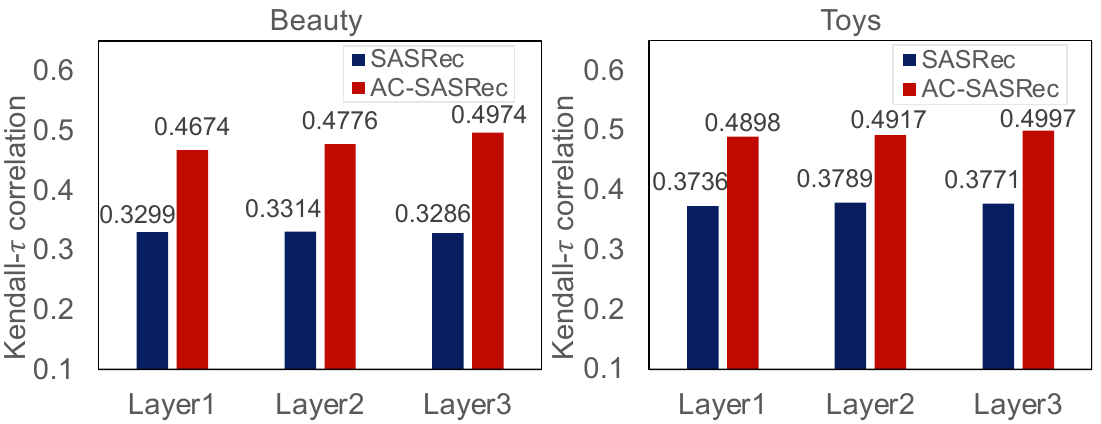}
  \caption{Comparison of the mean Kendall-$\tau$ correlation between attention weights and gradient importance measures. The results verify that our AC method can improve Kendall-$\tau$ correlation by a large margin.}
\label{fig:kendall-2}
\end{figure}
\begin{figure}[t]
  \centering
  \includegraphics[width=70mm]{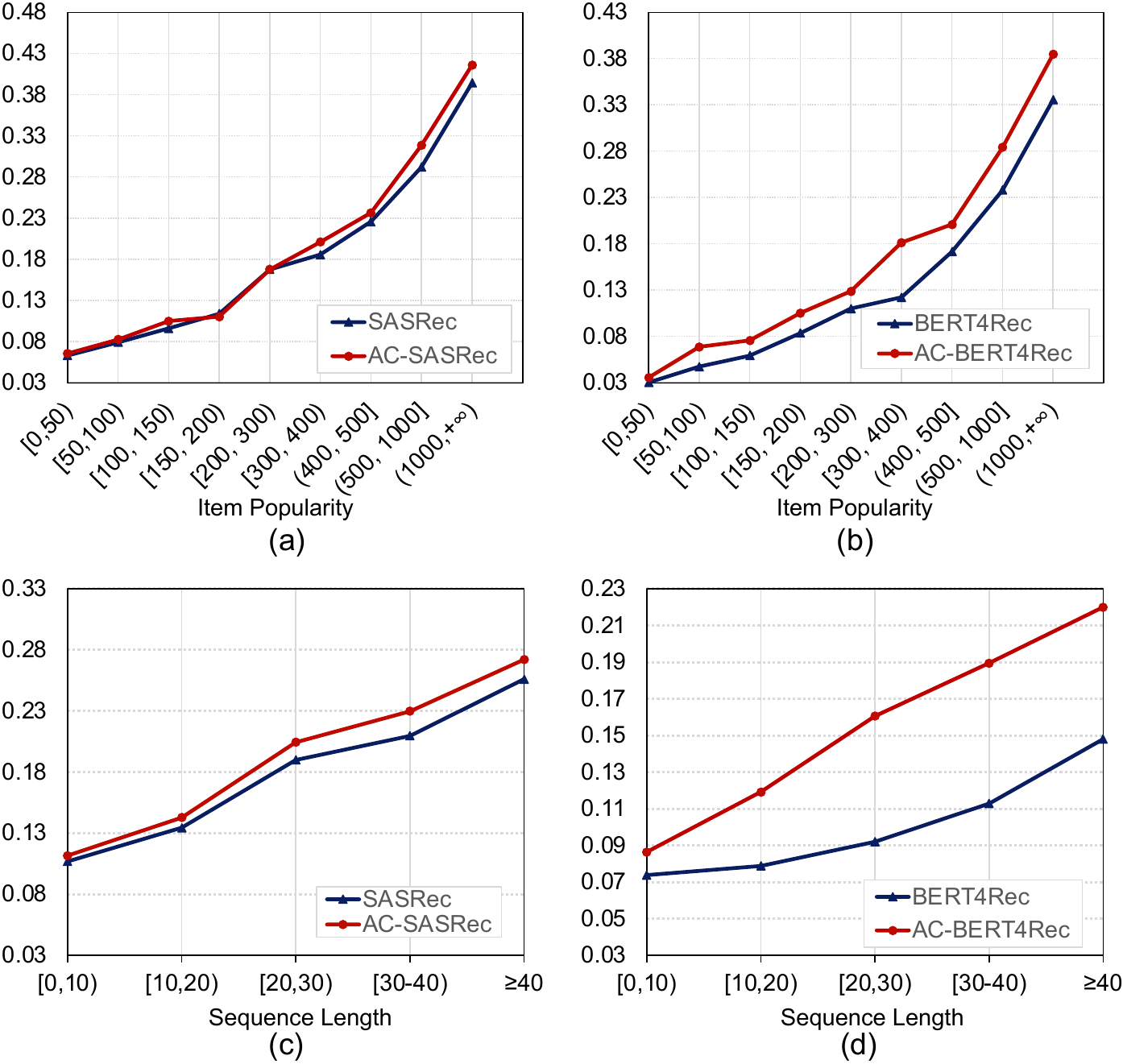}
    \caption{Performance comparison (Recall@20) between AC-TSR and TSR under different sequence lengths (\textit{i.e.}, number of training interactions of users) and item popularity (\textit{i.e.}, number of training interactions of items) on Amazon Beauty.}
\label{fig:item_len_pop}
\vspace{-10pt}
\end{figure}
\subsection{In-depth Analysis (RQ4)}
In this section, we further validate the effectiveness of our method from three different perspectives, including visualization, Kendall-$\tau$ correlations, and the performance comparison of AC-TSR across different sequence lengths and item popularity.

\noindent \textbf{Visualization and "erasing" experiment.} We visually compare the original attention weights with the calibrated ones generated by AC-TSR in Fig.~\ref{fig:intro}(b). It can be observed that the calibrated attention better aligns with the user's interests, resulting in improved recommendations. Moreover, Fig.~\ref{fig:intro}(a) indicates that removing the highest scoring items in the AC-TSR model noticeably degrades the model's performance on all the datasets, demonstrating that the learned attention weights in AC-TSR effectively aligns with the actual importance of items in the recommendation process.
 
\noindent \textbf{Correlation between attention weights and feature importance metrics.}
 Another way to examine whether attention mechanism focus on decisive inputs is computing the Kendall-$\tau$ correlation between attention weights and feature importance measures~\cite{jain2019attention}. Here we use gradient-based importance measures as feature importance indicators because they can effectively showcase feature significance with well-defined semantics~\cite{kindermans2019reliability}.
A lower value of Kendall-$\tau$ correlation indicates a higher inconsistency between attention weights and feature importance measures. This implies that items with high attention weights may not contribute significantly to the model's predictions~\cite{ross2017right}.
As shown in Fig.~\ref{fig:kendall-2}, for each layer, our AC-TSR approach significantly improves the correlation score, implying that attention calibration can help model focus on the decisive items.

\noindent \textbf{Performance w.r.t sequence lengths and item popularity.} We compare the original TSR model and our proposed AC-TSR model under varying sequence lengths and item popularity on the Amazon Beauty dataset, using Recall@20 as the performance metric. As shown in Fig~\ref{fig:item_len_pop}, the enhancements brought by AC-TSR over TSR are consistently observed across diverse sequence lengths and item popularity levels. We also note that the performance gains from Attention Calibration increase as the sequence length in the test set becomes longer. As for item popularity, the most prominent performance improvement is observed when the item popularity falls between 300 and 400.

\section{Conclusion and Future Work}
In this paper, we propose the AC-TSR framework, which can effectively calibrate the unreliable attention weights generated by existing transformer-based SR models. 
Specifically, AC-TSR adopts the spatial calibrator to substitute traditional positional embeddings, which directly utilizes low-level features including order and distance to yield position-aware attention weights.
Additionally, the adversarial calibrator is devised to adjust the attention weights according to the reassessed contribution of each historical item to the model prediction, making attention weights more robust to noisy input.
Comprehensive experiments on four benchmark SR datasets show that our approach outperforms competitive transformer-based SR methods, demonstrating the effectiveness of AC-TSR. 
In the future, we intend to investigate more lightweight calibrators and explore treating each calibrator as an adaptor to be incorporated into pre-trained transformer-based SR models, thereby achieving performance improvements with minimal computational expense.

\newpage
\balance
\bibliographystyle{ACM-Reference-Format}
\bibliography{sample-base}





\end{document}